\documentclass[aps,pre,showpacs,twocolumn,floats]{revtex4}
\usepackage{amssymb}

\usepackage{amsmath}
\usepackage{graphicx,psfig}
\usepackage{dcolumn}
\usepackage{bm}

\begin{document}

{\bf Reply to Comment on ``Universal Decoherence in Solids"}\\

The Comment of Openov \cite{Openev} on my Physical Review Letter
\cite{Letter} does not affect the universality of the decoherence
mechanism described in the Letter \cite{Letter}. The Letter
addressed the question of decoherence of quantum oscillations of a
two-state system in a solid when the oscillation frequency,
$\omega_0$, is below the Debye frequency, ${\omega}_D \sim
10^{13}\,$s$^{-1}$. I have shown that conservation laws (that is,
symmetry) mandate parameter-free interaction of the tunneling
variable with the phonon displacement field ${\bf u}$. For, e.g.,
a particle of mass ${\rm m}$, oscillating between degenerate
minima, ${\bf R} = \pm {\bf R}_0$, of a potential $U({\bf R})$,
conservation of the total linear momentum (particle + solid)
results in a decohering interaction
\begin{equation}\label{R}
{\rm m}{\dot{\bf R}}{\cdot}\dot{\bf u}\;.
\end{equation}
In quantum theory
\begin{equation}\label{com}
{\dot{\bf R}} = \frac{i}{\hbar}[{\cal{H}}, {\bf R}]\;,
\end{equation}
where ${\cal{H}}$ is the Hamiltonian of the system, while
\begin{equation}\label{Pi}
\dot{\bf u} = \frac{\cal{\bf \Pi}}{\rho} =   \frac{-i}{\sqrt{V}}
\sum_{{\bf k},i}\sqrt{\frac{{\hbar}{\omega}_{{\bf
k}i}}{2\rho}}\left( a_{{\bf k}i}{\rm e}^{i\bf k r} -
a^\dagger_{{\bf k}i}{\rm e}^{-i\bf k r} \right){\bf e}_{i}\;,
\end{equation}
where ${\cal{\bf \Pi}}$ is the momentum density of phonons,
canonically conjugate to ${\bf u}$. Here $V$ and $\rho$ are the
volume and the mass density of the system, $a^\dagger_{{\bf k}i}$
and $a_{{\bf k}i}$ are operators of creation and annihilation of
phonons, $\;{\omega}_{{\bf k}i}$ is the frequency of the phonon of
wave vector ${\bf k}$ and polarization $i$, and ${\bf e}_{i}$ are
unit vectors of polarization of the phonons.

For an orbital moment or a spin, ${\bf I}$, oscillating, e.g.,
between ${\pm I}$ projections onto a quantization axis,
conservation of the total angular momentum (${\bf I}$ + angular
momentum of a solid) results in a decohering interaction
\begin{equation}\label{I}
 \frac{1}{2}\,{\bf I}{\cdot}({\bm
\nabla}{\times}\dot{\bf u})\, ,
\end{equation}
with $\dot{\bf u}$ given by Eq.\ (\ref{Pi}).

Equations (\ref{R}) and (\ref{I}) were used in my Letter
\cite{Letter} to obtain universal lower bound on the decoherence
of solid-state qubits. Parameter-free decoherence rates were
computed within perturbation theory, making use of the Fermi
golden rule. It gives accurate results for, e.g., tunneling of
electron or tunneling of a spin, and applies to all solid-state
systems that fit definition of a qubit. At low temperature, $k_BT
\ll \hbar \omega_0$, the decoherence occurs due to spontaneous
resonant emission of a phonon of frequency $\omega_0$. Openov
observed \cite{Openev} that for a heavy particle oscillating at a
high frequency, the decoherence due to Eq.\ (\ref{R}) may become
so strong that standard perturbation theory and the Fermi golden
rule may no longer apply. I would like to stress that this
observation in no way affects the universality of the mechanism of
decoherence pointed out in my Letter. Parameter-free equations
(\ref{R}) and (\ref{I}) describe unavoidable decohering effect of
the elastic environment, which is mandated by the conservation
laws regardless of the strength of decoherence. They apply to all
problems of quantum tunneling of a particle in a solid
\cite{Letter}, to transitions between quantum states of a spin in
a crystal field \cite{Letter,C,CM}, and to tunneling between
macroscopic quantum states of a superconducting current \cite{CK}.
\\
\\
E. M. Chudnovsky\\
Physics Department, CUNY Lehman College\\
Bronx, NY 10468-1589

%

\begin{thebibliography}{99}

\bibitem{Openev} {L. A. Openov}, Phys. Rev. Lett. {\bf 93}, 158901
(2004); cond-mat/0410106.

\bibitem{Letter}
{E. M. Chudnovsky}, Phys. Rev. Lett. {\bf 92}, 120405 (2004).

\bibitem{C}
{E. M. Chudnovsky}, Phys. Rev. Lett. {\bf 72}, 3433 (1994).

\bibitem{CM}
{E. M. Chudnovsky and X. Martinez-Hidalgo}, Phys. Rev. {\bf B66},
054412 (2002).

\bibitem{CK}
{E. M. Chudnovsky and A. B. Kuklov}, Phys. Rev. {\bf B67}, 064515
(2003).

\end{thebibliography}
\end{document}